\documentclass[12pt,twocolumn]{article}
\usepackage{amsmath,amssymb}
\usepackage{epsfig}

\title{Interaction-Free Measurement Applied To Quantum Computation: A New ``cnot'' Gate.}
\author{A. A. M\'{e}thot$^\dag$\thanks{Present address: D\'epartement d'informatique et de recherche op\'erationnelle, Universit\'e de Montr\'eal, CP 6128 succ Centre-Ville, Montr\'eal QC, H3C 3J7, Canada}\, and Kai Wicker$^\S$\\
\\
$^\dag$Institut f\"{u}r Theoretische Physik\\
Universit\"{a}t Heidelberg\\
Philosophenweg 16, D-69120 Heidelberg\\
Deutschland\\
methotan@iro.umontreal.ca\\
\\
$^\S$Fakult\"{a}t f\"{u}r Physik und Astronomie\\
Universit\"{a}t Heidelberg\\
Albert-Ueberle-Str. 11, D-69120 Heidelberg\\
Deutschland\\
kwicker@ix.urz.uni-heidelberg.de}
\date{\today}

\begin{document}
%%%%%%%%%%%%%%%%%%%%%%%%%%%%%%%%%%%%%%%%%%%%%%%%%%
\begin{titlepage}
\maketitle
\thispagestyle{empty}
\begin{abstract}
In this paper, we propose a new direction of research for the realization of the quantum controlled-not gate based on a technique called ``interaction-free measurement'', where qubits are two-level atoms (or ions) and information is mediated from one qubit to another by lasers in superpositions of $\pi$ and $2\pi$ pulses. We investigate the advantages and limitations of such a gate and discuss possible applicability.
\end{abstract}
\end{titlepage}
\begin{figure}[!htb]
\input{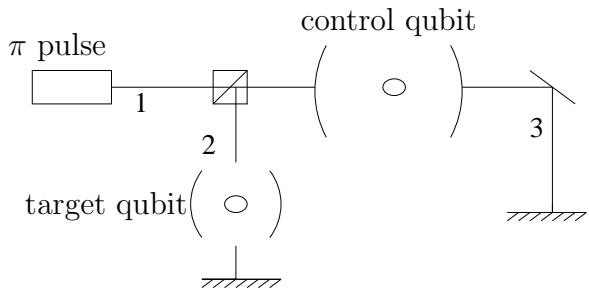}
\caption{\label{dessin2}Schematic view of the cnot gate}
\end{figure}
\begin{figure}[!htb]
\input{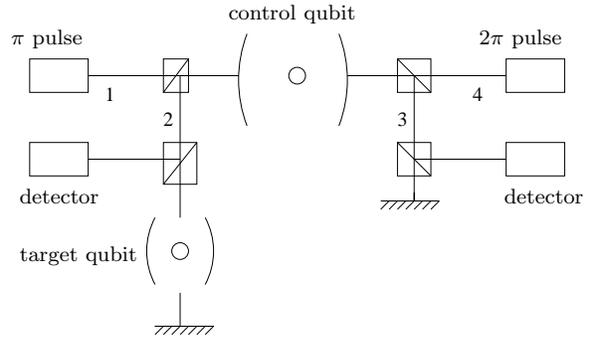}
\caption{\label{dessin}Schematic view of the improved cnot gate}
\end{figure}

%\pagenumbering{arabic}
%%%%%%%%%%%%%%%%%%%%%%%%%%%%%%%%%%%%%%%%%%%%%%%%%%
In the recent past, physicists and computer scientists alike have turned their interest to quantum information processing. Since the quantum computer seems so promising on paper, with discoveries like super-fast algorithms as Shor's algorithm \cite{shor}, many have attempted to create a vehicle to process quantum information \cite{cirac,turchette,gershenfeld,zanardi,shnirman,loss}. It is well known that one model of computation is the circuit model, in which every computer is made out of logic gates and that every logic gate can be made out of an appropriate combination of universal gates. For classical com\-puters, one needs a ternary gate (like the Tiffoli gate) to have a universal gate, while quantum computers only need the controlled-not gate (called cnot gate henceforth), sometime called controlled-inversion gate or xor gate, and proper unary gates ($SU(2)$ group). The classical reversible cnot gate is the logic operation on two bits that takes $(a,b)\rightarrow (a,b\oplus a)$.\\
\indent
The generalization to qubits and quantum cnot gates is obvious by linearity of the Hilbert space. Therefore most attempts to build a quantum computer so far consist of creating a working and efficient quantum cnot gate. To date, there exist at least six se\-rious proposals for universal quantum computing: ion traps \cite{cirac}, cavity quantum electrodynamics (cavity QED)\cite{turchette}, li\-quid nuclear magnetic resonance (NMR) \cite{gershenfeld}, super-conducting quantum interference devices (SQUIDs) \cite{zanardi,shnirman}, quantum dots \cite{loss} and linear optics \cite{knill}.\\
\indent
In order to have a working, successful quantum computer, one needs the quantum computing device to have a long time storage capacity for qubits (long relative decoherence time), a good isolation of the system from the environment, an efficient way to read-out the information, an initial state preparation system and a precise and universal set of gates \cite{divincenzo,preskill,brassard}. Although the ion traps scheme and the NMR scheme seem to be the most promising technics so far, they both fail to meet all these requirements \cite{cirac,gershenfeld,preskill,hughes,steane,warren}. While both may be relatively slow (NMR has a theoretical speed limit of $10Hz$) and have initialization and scaling problems (fundamental issues limits the NMR computer to the order of 100 qubits), NMR also has read-out problems \cite{hughes,warren}. Although slowness is not really a big disadvantage when one compares quantum computers to classical computers because of the super-fast algorithms, one can still hope for a better quantum scheme, a scheme that can be faster and still leave some room for improvement. Unfortunately scaling is an essential criterion for the creation of a quantum computer, in order to solve problems involving more qubits. As for cavity QED, the scheme suffers from a high rate of decoherence, since the cavity has a high rate of decay, and has non-trivial technological gates \cite{hughes,pellizzari,liu}. The linear optics quantum computation scheme is also technolo\-gically non-trivial. It would require a single photon source to work at the highest level of efficiency \cite{knill,knill2}. SQUIDs and quantum dots also have high rates of decoherence and they lack a coherent gate yet \cite{hughes,brown}, although some claim they have observed coherence recently in SQUIDs \cite{friedman}.\\
\indent
Since these problems are technologically unresolved yet, although some show early signs of improvement, further improvements are required in order for them to yield better results. One also needs to investigate other possibilities. In this paper, we propose a new direction in the field, that has its own advantages and drawbacks. We propose the use of Interaction-Free Measurement (IFM) to carry the information from one qubit to another and show how this can be done efficiently.\\
\indent
As pointed out by many, IFM are not truly interaction free, but would be better described by ``energy-exchange-free interactions'' \cite{karlsson,pavicic}. IFM are a way of \-measuring, usually by means of an interferometer, where the probed object has only a small pro\-bability of absorbing energy \cite{karlsson,pavicic,vaidman,potting,kwiat}. When applied to the measurement of the state of a quantum system, the probe becomes entangled with the probed object, with only a small probability of irreversibly losing the quantum information. The information is simply transferred from one system to non-local correlations. No information is gained on the state of the object and the wave-function does not collapse until one performs a read-out on the probe.\\
\indent
IFM could be used in the creation of quantum cnot gates. First, one probes a two-level quantum object, referred to as control qubit, by means of an IFM. Since no measurement is made, the quantum state will not collapse and all the information it contains is preserved. The probe is now entangled with the control qubit. Let's assume that the probing system is set to send a signal $A$ if the control qubit is in the ground state $\left| 0 \right>$ and a signal $B$ if it is in the excited state $\left| 1 \right>$.\\
\begin{equation}
\begin{split}
&\left| 0 \right> \rightarrow \left| 0 \right> \left| A \right>\\
&\left| 1 \right> \rightarrow \left| 1 \right> \left| B \right>
\end{split}
\end{equation}
\indent
Event $A$ and event $B$ must be mutually exclusive ($\left < A | B \right > = 0 $) and can be paths, frequencies, polarizations or anything else depending on the IFM scheme. Now we place the second qubit in a way that it can only interact with signal $B$. Therefore, if signal $B$ can alter the state of the second qubit $\left| \Psi \right>$ ($\left| B \right> \left| \Psi \right> \rightarrow \left| \Psi_{orthogonal} \right>$, while $\left| A \right> \left| \Psi \right> \rightarrow \left| \Psi \right>$), the only step left in order to have a working cnot gate is to reset the probe to a standard state independent of $\left| \Psi \right>$. An advantage that can be readily seen is that the scaling to quantum computer seems promising, one only has to place the second qubit in another IFM device. There is no fundamental limit on the number of qubits that could be involved in a calculation. The only limits are on technology, budget or available space.\\
\indent
This can be done realistically with an IFM using a Fabry-Perot interferometer \cite{karlsson}. One places a two-level ion (or atom) in an ion (atom) trap in the middle of the Fabry-Perot interferometer. The length of the interfero\-meter is set such that it will have maximum transmitivity ($T=1$) when the ion (atom) is in the ground state, see Figure~\ref{dessin2}. Then one sends a laser $\pi$ pulse into one end of the interferometer. If the ion (atom) is in its ground state the light will be transmitted through the cavity, while if it is in the excited state the light will be reflected onto a second qubit, called target qubit, with which it will interact. The entanglement between the Fabry-Perot interferometer and light has been realized by the Karlsson-Bj\"ork-Forsberg group \cite{karlsson} for imaging purposes. The reflectivity of the interferometer is limited only by the quality of the mirror and can be made arbitrarily close to one ($R \approx 1$). The initial state of the control qubit can be modified with lasers ($q \pi$ pulse where $q$ is a real number) sent onto the ion (atom), perpendicularly to the axe of the cavity, thus preparing the initial state in the wanted superposition.\\
\indent
Of course, as Renninger pointed out \cite{renninger}, the non observation of the result is a meas\-ure\-ment. So in order to prevent the wave-function from collapsing, one has to delete the information in a coherent way. To prevent gaining information on which way the light went, one can put mirrors at the end of each trajectory to reflect the light back into the laser. Every mirror in the circuit will induce a phase shift on the pulse, but these phases don't matter in our cases, they don't modify the interactions.
\\
\indent
Let's assume the control qubit is in the state $\alpha \left| 0 \right> + \beta \left| 1 \right>$ and the target qubit in the state $\gamma \left| + \right> + \delta \left| - \right>$ (to avoid confusions). The system can be formally described by noting the $\pi$ pulse $\left| \pi \right>$ and appending a subscript corresponding to which arm the pulse is in, see Figure~\ref{dessin2}. The initial state, when the pulse is emitted, can be described as:
\begin{equation}
\label{begin}
\left| \pi_1 \right> \left(\alpha \left| 0 \right> + \beta \left| 1 \right> \right) \left( \gamma \left| + \right> + \delta \left| - \right> \right)
\end{equation}
\indent
Then the $\pi$ pulse hits the Fabry-Perot cavity and becomes entangled with the control qubit:
\begin{equation}
\begin{split}
\label{super}
\Rightarrow \left( \alpha \left| \pi_3 \right> \left| 0 \right> + \beta \left| \pi_2 \right> \left| 1 \right> \right)\\
\otimes \left( \gamma \left| + \right> + \delta \left| - \right> \right)
\end{split}
\end{equation}
\indent
The part of the $\pi$ pulse that is in the second arm interacts with the second qubit, hence changing its state to:
\begin{equation}
\begin{split}
\Rightarrow \alpha \gamma \left| \pi_3 \right> \left| 0 \right> \left| + \right> + \alpha \delta \left| \pi_3 \right> \left| 0 \right> \left| - \right>\\
+ \beta \gamma \left| \pi_2 \right> \left| 1 \right> \left| - \right> + \beta \delta \left| \pi_2 \right> \left| 1 \right> \left| + \right>
\end{split}
\end{equation}
\indent
Then the light is reflected back to the cavity where the initial entanglement is undone:
\begin{equation}
\begin{split}
\label{end}
\Rightarrow \left| \pi_1 \right> ( &\alpha \left| 0 \right> ( \gamma \left| + \right> + \delta \left| - \right> )\\
+ &\beta \left| 1 \right> ( \gamma \left| - \right> + \delta \left| + \right> ))
\end{split}
\end{equation}
\indent
As can be seen by comparing~\eqref{begin} and~\eqref{end}, a cnot operation has been performed.\\
\indent
Since a $\pi$ pulse is made of a macroscopic number of photons, a superposition of paths as in~\eqref{super} is extremely fragile. The decoherence of a single photon is enough for the superposition to collapse to a single path.\\
\indent
One can circumvent this situation by adding a $2\pi$ pulse at the other end of the interferometer, see Figure~\ref{dessin}. By adding this new pulse, one prevents gaining ``which way'' information on the system by absorption of a single photon, by a mirror or any other part of the system, since the photons from both lasers are indistinguishable. The number of photons in each pulse can also be made arbitrarily close to one another by choosing a $2n\pi$ pulse and a $(2n+1)\pi$ pulse, where $n$ is an integer large enough. Thus, one cannot distinguished which pulse went through which path by calculating the recoil of the end mirrors or the energy absorbed by them. The two pulses are also indistinguishable by counting the number of photons in them, since both lasers are a superposition of number states \cite{fuchs}. The extra pulse will not disturb the system, since a $2\pi$ pulse leave the state of the targeted ion (atom) unchanged. Let us examine what happens in such a gate in details. In the initial state, both pulses are emitted:
\begin{equation}
\left| \pi_1 \right> \left| 2\pi_4 \right> \left( \alpha \left| 0 \right> + \beta \left| 1 \right> \right) \left( \gamma \left| + \right> + \delta \left| - \right> \right)
\end{equation}
\indent
They both interact with the control qubit, and both become entangled:
\begin{equation}
\begin{split}
\Rightarrow \left( \alpha \left| \pi_3 \right> \left| 2\pi_2 \right> \left| 0 \right> + \beta \left| \pi_2 \right> \left| 2\pi_3 \right> \left| 1 \right> \right)\\
\cdot \left( \gamma \left| + \right> + \delta \left| - \right> \right)
\end{split}
\end{equation}
\indent
The part of the $\pi$ pulse that is in the second arm will interact with the target qubit while the $2\pi$ pulse does not affect the state of the qubit:
\begin{equation}
\begin{split}
\Rightarrow \alpha \gamma \left| \pi_3 \right> \left| 2\pi_2 \right> \left| 0 \right> \left| + \right>\\
+ \alpha \delta \left| \pi_3 \right> \left| 2\pi_2 \right> \left| 0 \right> \left| - \right>\\
+ \beta \gamma \left| \pi_2 \right> \left| 2\pi_3 \right> \left| 1 \right> \left| - \right>\\
+ \beta \delta \left| \pi_2 \right> \left| 2\pi_3 \right> \left| 1 \right> \left| + \right>
\end{split}
\end{equation}
\indent
Coming back through the interferometer, the pulses become unentangled:
\begin{equation}
\begin{split}
\Rightarrow \left| \pi_1 \right> \left| 2\pi_4 \right> ( &\alpha \left| 0 \right> ( \gamma \left| + \right> + \delta \left| - \right> )\\
+ &\beta \left| 1 \right> ( \gamma \left| - \right> + \delta \left| + \right> ))
\end{split}
\end{equation}
\indent
In case one wants to read out the information, one simply adds detectors in each trajectory that can be turned on at any desired time, see Figure~\ref{dessin}. One simply sends a pulse and sees in which detector it ends up.
\\
\indent
While this scheme seems to require some fine technological tuning (as all the other schemes) and to require ample room, it does seem to have its own advantages. First, the only obstacle to scaling is technological, no fundamental limitations are imposed. The read-out is also technologically feasible. Relaxation time should, in principle, be, at least as long as the ion trap scheme. The low decoherence rate by the gate is insured by the use of a $2\pi$ pulse, with the $\pi$ pulse, which prevents gaining ``which way'' information. The pulses are also intrinsically stable to decoherence. A missing photon from one of the pulses will not affect the system very much, since the pulse will remain essentially a $\pi$ or $2\pi$, errors may occur infrequently. In fact, a missing photon from a $n\pi$ pulse will have the effect of rotating the pulse to a $(n-\epsilon)\pi$ pulse, where $\epsilon$ is small. Such a small error can be corrected efficiently by the various quantum correction schemes \cite{steaneerr,shorerr}. It also seems that this gate could be faster then the NMR or cavity QED schemes. Speed is only limited by the time a pulse is traveling through the apparatus and the time it takes to change the state of an ion (atom), which lies in the order of kHz (or faster). \\
\indent
This scheme is in some ways similar to the one proposed by S. Haroche \cite{haroche1,haroche2}, where he and his group uses a Rydberg atom to maximally entangle two cavities together. The problem is that they cannot undo the entanglement between the cavities and the atom, therefore losing all coherence when the atom interacts with the environment.
\\
\indent
In conclusion, there exist many good ideas for quantum computers, but it does not mean we should to stop looking for other solutions. In this paper we gave a detailed suggestion of a new idea that seems to have its own advantages and disadvantages over the other proposed schemes that exist so far. The idea is to use IFM to mediate the information from one qubit to another. We gave a detailed description on how this could be realized using a Fabry-Perot interferometer. Other similar schemes are possible, like one using the quantum Zeno effect \cite{kwiat}.\\
\indent
We would like to thank the University of Heidelberg where the major part of this work was realized. We also would like to thank G. Brassard, F. Bussi\`eres and D. Poulin for helpful discussions. This work was supported by the Institute for Theoretical Physics of Heidelberg.
%%%%%%%%%%%%%%%%%%%%%%%%%%%%%%%%%%%%%%%%%%%%%%%%%%


\begin{thebibliography}{50}
\bibitem{shor}P. W. Shor, SIAM J. Computing \b{26}, 1484 (1997).
\bibitem{cirac}J. I. Cirac and P. Zoller, Phys. Rev. Lett. \textbf{74}, 4091 (1995).
\bibitem{turchette}Q. A. Turchette, C. J. Hood, W. Lange, H. Mabuchi and H. J. Kimble, Phys. Rev. Lett. \textbf{75}, 4710 (1995).
\bibitem{gershenfeld}N. A. Gershenfeld and I. L. Chuang, Science \textbf{275}, 350 (1997).
\bibitem{zanardi}P. Zanardi and M. Rasetti, Phys. Letts. A \textbf{264}, 94 (1999).
\bibitem{shnirman}A. Shnirman, G. Sch\"on and Z. Hermon, Phys. Rev. Lett. \textbf{79}, 2371 (1997).
\bibitem{loss}D. Loss and D. P. DiVincenzo, Phys. Rev. A \textbf{57}, 120 (1998).
\bibitem{knill}E. Knill, R. Laflamme and G. Milburn, quant-ph/0006088.
\bibitem{divincenzo}D. P. DiVincenzo, quant-ph/0002077.
\bibitem{preskill}J. Preskill, preprint.
\bibitem{brassard}G. Brassard, preprint.
\bibitem{hughes}R. J. Hughes \textit{et al.}, Fortsch. Phys. \textbf{46}, 329 (1998).
\bibitem{steane}A. M. Steane and D. M. Lucas, quant-ph/0004053.
\bibitem{warren}W. S. Warren, Science \textbf{277}, 1688 (1997).
\bibitem{pellizzari}T. Pellizzari, S. A. Gardiner, J. I. Cirac and P. Zoller, Phys. Rev. Lett. \textbf{75}, 3788 (1995).
\bibitem{liu}X. M. Liu, M. Hug and G. J. Milburn, quant-ph/0001056.
\bibitem{knill2}E. Knill, R. Laflamme and G. Milburn, quant-ph/0006120.
\bibitem{brown}K. R. Brown, D. A. Lidar and K. B. Whaley, quant-ph/0105102.
\bibitem{friedman}J. R. Friedman, V. Patel, W. Chen, S. K. Tolpygo and J. E. Lukens, nature \textbf{406}, 43 (2000).
\bibitem{karlsson}A. Karlsson, G. Bj\"ork and E. Forsberg, Phys. Rev. Lett. \textbf{80}, 1198 (1998).
\bibitem{pavicic}M. Pavicic, Phys. Lett. A \textbf{223}, 241 (1996).
\bibitem{vaidman}A. Elitzur and L. Vaidman, Foundation of physics \textbf(23), 987 (1993).
\bibitem{potting}S. P\"otting, E. S. Lee, W. Schmitt, I. Rumyantsev, B. Mohring and P. Meystre, Phys. Rev. A \textbf{62} 060101(2000).
\bibitem{kwiat}P. G. Kwiat, A. G. White, J. R. Mitchell, O. Nairz. G. Weihs, H. Weinfurter and A. Zeillinger, Phys. Rev. Lett. \textbf{83}, 4725 (1999).
\bibitem{renninger}M. Renninger, Z. Phys. \textbf{158}, 417 (1960).
\bibitem{fuchs}S. J. van Enk and C. A. Fuchs, quant-ph/0104036.
\bibitem{steaneerr}A. M. Steane, Phys. Rev. Lett. \textbf{77}, 793 (1995).
\bibitem{shorerr}A. R. Calderbank and P. W. Shor, Phys. Rev. A \textbf{54}, 1098 (1996).
\bibitem{haroche1}A. Rauschenbeutel, P Bertet, S Osnaghi, G. Nogues, M. Brune, J. M. Raimond and S. Haroche, quant-ph/0105062.
\bibitem{haroche2}S. Osnaghi, P. Bertet, A. Auffeves, P. Maioli, M. Brune, J. M. Raimond and S. Haroche, quant-ph/0105063.
\end{thebibliography}
\end{document}